# Discovery of water vapor around IRC+10216 as evidence for comets orbiting another star

Gary J. Melnick*, David A. Neufeld†, K.E. Saavik Ford†, David J. Hollenbach‡ & Matthew L.N. Ashby*

 *Harvard-Smithsonian Center for Astrophysics, 60 Garden Street, Cambridge, MA 02138, USA.

† Department of Physics & Astronomy, The Johns Hopkins University, 3400 N. Charles Street, Baltimore, MD 21218, USA.

‡ NASA/Ames Research Center, Moffett Field, CA 94035, USA.

Since 1995, astronomers have discovered planets with masses comparable to that of Jupiter (318 times Earth's mass) in orbit around approximately 60 stars[1]. Although unseen directly, the presence of these planets is inferred by the small reflex motions that they gravitationally induce on the star they orbit; these result in small periodic wavelength shifts in the stellar spectrum. Since this method favors the detection of massive objects orbiting in close proximity to the star, the question of whether these systems also contain analogs of the smaller constituents of our Solar System has remained unanswered. Using an alternative approach, we report here observations of an aging carbon-star, IRC+10216, that reveal the presence of circumstellar water vapor, a molecule not expected in measurable abundances around such a star and thus a distinctive signature of an orbiting cometary system. The only plausible explanation for this water vapor is that the recent evolution of IRC+10216, – which is accompanied by a prodigious increase in its luminosity – is now causing the **vaporization of a collection of orbiting icy bodies, a process first considered in a previous theoretical study[2].**

The notion that large numbers of icy bodies may orbit other stars has its parallel in the Solar System's Kuiper Belt. The Kuiper Belt is a collection of cometary nuclei, located roughly in the plane of the ecliptic and beyond the orbit of Neptune, beginning at approximately 30 astronomical units (AU; 1 AU = 1.5 x $10^{13}$ cm) from the Sun, and thought to extend to a few hundred AU from the Sun. About $10^5$ Kuiper Belt objects exist with diameters larger than 100 km – the largest being Pluto (~2400 km) – and it is likely that there is a significantly greater number of smaller objects. From their orbital characteristics, it seems very probable that the Kuiper Belt is the present-day source of short-period comets. Long-period comets are presumed to originate from the Oort Cloud, an approximately spherically symmetric cloud of cometary nuclei located well beyond the Kuiper Belt at distances from the Sun of between 3000 and $10^5$ AU. For main sequence stars, such as the Sun, the stellar radiation field is sufficiently dilute beyond about 3 AU that icy bodies are unaffected. Even stars possessing masses between 1.5 and 4 times that of the Sun, the range which is believed to encompass the mass of IRC+10216, would have lacked the luminosity to have vaporized icy bodies beyond about 40 AU during their main sequence phase.

After exhausting the supply of hydrogen and helium in its core, a star of a few solar masses evolves off the main sequence and begins burning hydrogen and helium in thick shells surrounding a core enriched in carbon. By the time the star has reached this so-called asymptotic giant branch (AGB) phase, the stellar radius has increased by a factor of several hundred to a thousand – in the case of IRC+10216 reaching a value of ~ 5 AU (the radius of Jupiter's orbit) – while the luminosity has increased by a factor of between 100 and 3000. It has been pointed out[2] that this luminosity increase is sufficient to cause the vaporization of any icy body orbiting within several hundred AU, thus resulting in the release of significant amounts of water vapor provided the star is surrounded by a Kuiper Belt analog. Objects orbiting in an analog of the Oort Cloud, however, would remain unaffected.

Unfortunately, the detection of water vapor toward an evolved star does not unambiguously constitute evidence for the presence of Kuiper Belt-type objects. For those cases in which the circumstellar gas is oxygen-rich, it is predicted that almost all of the carbon will combine with oxygen to form CO, with much of the remaining oxygen reacting to form $H_2O$. Strong water vapor emission is expected from such oxygen-rich stars and, indeed, this is observed (e.g., refs. 3, 4). During the final phase of their evolution, however, convection of material from the core can enhance the outer layers of AGB stars with carbon, altering the carbon-to-oxygen ratio in the outflowing gas. For those AGB stars, like IRC+10216, possessing circumstellar envelopes in which carbon is the most abundant heavy element, it is predicted – by contrast – that the equilibrium chemistry will drive all of the oxygen into CO with little remaining to form other molecules[5]. Thus, the detection of water vapor toward a carbon-rich AGB star raises the possibility that icy bodies are being vaporized, since the carbon-rich outflow from the star has no chemical route to produce significant water vapor.

Of the more than 100 known carbon-rich circumstellar envelopes, IRC+10216 is by far the brightest and most intensively studied. A search was conducted for water vapor surrounding IRC+10216 in the rotational ground-state $1_{10}$-$1_{01}$ 556.936 GHz line of ortho-$H_2{}^{16}O$ using NASA's *Submillimeter Wave Astronomy Satellite* (*SWAS*) during the periods 1999 May 11–31; 1999 November 12 – December 12; 1999 December 24 – 2000 January 8; and 2000 February 5 – March 9. The spectrum obtained is shown in Fig. 1. The line profile is well fit by the parabolic curve expected for optically thick emission from an unresolved, constant velocity outflow. The integrated antenna temperature is 0.39 K km s$^{-1}$, uncorrected for the *SWAS* aperture efficiency of 0.66. The line width at half maximum is 25.0 km s$^{-1}$ and the LSR line center is -23.5 km s$^{-1}$, in good agreement with the dynamical quantities measured in the lines of other species toward this source (e.g., ref. 7).

An identification other than the $1_{10}$-$1_{01}$ water vapor line can be ruled out. Standard line catalogues[8], although admittedly incomplete, reveal no molecules possessing both a transition with a frequency within a few km s$^{-1}$ of 556.936 GHz and an upper level fractional population that is plausibly high enough to account for the observed line strength. Furthermore, due to its low excitation energy and to the small partition function of the water molecule, the $1_{10}$-$1_{01}$ water transition is excited very effectively and thus the abundance required of any other molecule responsible for the observed feature would have to be considerably greater than that derived below for water. The only plausible molecule of sufficient abundance is HCN, for which a complete line catalogue[9] reveals no transition of the required frequency in any vibrational state of energy lower than 15,000 K.

The water abundance is derived by modeling the circumstellar envelope as an outflow with the properties given in Table 1. Over the range of mass-loss rates inferred previously (refs. 12,13) for IRC+10216, $\dot{M} = 2 - 5$ x $10^{-5}$ M$_\odot$ yr$^{-1}$, we find that the $1_{10}$ $-1_{01}$ line luminosity can be expressed as

L($1_{10} - 1_{01}$) = 3.6 x $10^{21}$ ( $\dot{M}$ /$10^{-5}$ M$_\odot$ yr$^{-1}$) (x[H$_2$O]/$10^{-7}$)$^{0.5}$ W

where x[H$_2$O] is the water abundance relative to H$_2$. The *SWAS*-measured $1_{10}$-$1_{01}$ line flux is 1.03 x $10^{-20}$ W cm$^{-2}$, corresponding to a luminosity of 3.6 x $10^{22}$ W for an isotropic source at an assumed distance of 170 pc (ref. 12). The implied water abundance is

x[H$_2$O] = 1.1 x $10^{-6}$ ( $\dot{M}$ /3 x $10^{-5}$ M$_\odot$ yr$^{-1}$)$^{-2.0}$ =  4 − 24 x $10^{-7}$ .

The derived water abundance in IRC+10216 can be compared with the abundance that would be expected if orbiting icy bodies are not present. In thermochemical equilibrium (TE), the initial water abundance in the outflowing gas is expected to be only $\sim 10^{-12}$ (ref. 14). The chemical composition of the gas can subsequently be altered by shocks[14] in the inner part of the circumstellar envelope – a process invoked previously to explain observations of water vapor toward the protoplanetary nebula CRL 618 (ref. 15) – or by photodissociation in the outer envelope[5]. The best current models[5,14] for these two processes, constructed specifically for the source IRC+10216, indicate (1) that pulsationally-driven shocks – if present – would actually reduce the water abundance below its already small TE value[14]; and (2) that photodissociation of CO followed by incorporation of O into $H_2O$ can indeed increase the water vapor abundance, but only to a value $\sim 5 \times 10^{-12}$ (Millar and Herbst 2001, personal communication). Thus, apart from the evaporation of icy bodies, there is no known mechanism that comes within a factor of $10^4$ of explaining the water abundance observed in IRC+10216.

We now turn to the evaporation scenario, in which the evaporation of orbiting icy bodies is driven by the increasing luminosity of IRC+10216. The stellar evolution is characterized by a rise in luminosity, modulated by periodic stellar thermal pulsations. As the luminosity of the star increases, a zone of evaporation moves outwards through the orbiting material (Fig. 2). The inner edge of the evaporation zone is defined by the distance within which the largest icy bodies have already been entirely vaporized, and the outer edge is defined by the distance at which the icy bodies are warmed above the sublimation temperature of water ice. By the time the convection of carbon has made the photospheric C/O ratio greater than 1, it is found that for distances up to 75 AU even icy bodies as large as Pluto will have been completely vaporized (Ford and Neufeld, 2001, in preparation); thus, in the evaporation scenario, the release of water vapor currently observed occurs beyond about 75 AU. The evaporation is a continuous process in which the ongoing vaporization of icy bodies replenishes the water vapor as it flows through the outflow and is ultimately photodissociated by the interstellar ultraviolet radiation field.  New icy bodies continuously replenish the sublimated objects because the evaporation zone  steadily moves outwards as the stellar luminosity increases.

Detailed models for the evaporation of icy bodies have been constructed by Ford and Neufeld (2001, in preparation), but a simple argument reproduces the essential result.  The current water mass loss rate is given by

$$\dot{M} \; x(H_2O) \; [\mu(H_2O)/\mu(H_2)] \sim 3 \; x \; 10^{-10} \; M_\odot \; yr^{-1}$$

where $[\mu(H_2O)/\mu(H_2)] = 9$ is the ratio of the molecular mass of water to that of $H_2$.  As the current rate of hydrogen mass-loss can have been sustained for only $\sim 10^5$ yr, a total ice mass of  $\sim 3 \; x \; 10^{-5} \; M_\odot \sim 10$ Earth masses is sufficient to supply the necessary water vapor.  This is comparable to the original mass of water-ice believed present in the Solar System's Kuiper Belt (ref. 16).

The range of radii over which water vapor is present in the outflow, and therefore the range of excitation conditions, determines the emergent water spectrum.  A complete 2.4 – 197 micron spectrum of IRC+10216, obtained by the European Space Agency's (ESA) *Infrared Space Observatory (ISO)*[17,18], allows upper limits to be placed on a number of water transitions.  Based upon our line flux predictions, the most easily detectable $H_2O$ line in this wavelength region should be the $2_{12} - 1_{01}$ 179.53 micron transition.  Unfortunately, because this line is blended with a stronger HCN line at 179.51 microns, a flux determination in the water line is difficult.  The next most easily detectable water line in this band is predicted to be the $3_{03} - 2_{12}$ transition at 174.63 microns, which is free from overlap with other lines.  The upper limit on the *ISO*-measured strength of the $H_2O$ $3_{03} - 2_{12}$ line is marginally below what would be expected were the water abundance uniform throughout the circumstellar envelope, but is entirely consistent with our model in which the water emission arises from the cooler (T~100K), lower density gas beyond 75 AU where icy bodies are evaporating.

We are grateful to A. Glassgold for his valuable help and insights. We thank Tom Millar and Eric Herbst for communicating unpublished results from their chemical model of IRC+10216. This work was supported by NASA.

Table 1. The physical conditions assumed in our model for the circumstellar outflow of IRC+10216 are listed. Our model yields predictions for the luminosities of several water transitions as a function of water abundance, using two independent methods that are found to be in excellent agreement with each other: a fast escape probability method based upon the assumption that the large velocity gradient (LVG) approximation applies; and a more exact Monte Carlo method[10,11] that is much more expensive numerically. The effects of radiative pumping of rotational transitions by the far-infrared radiation field are included; vibrational excitation effects are negligible.

| **Model Assumptions**: | | **Ref.** |
|---|---|---|
| Stellar radius | $7.65 \times 10^{13}$ cm | 12 |
| Photospheric temperature | 2010 K | 12 |
| Dust shell inner radius, $R_i$ | $2.58 \times 10^{14}$ cm | 12 |
| Dust shell outer radius | $1 \times 10^{17}$ cm | 12 |
| Outflow velocity, $v_r$ | $14.5 \, (1.00 - 0.95[R_i/R])^{0.5}$ km s$^{-1}$ | 12 |
| Turbulent velocity | 0.65 km s$^{-1}$ | 12 |
| $H_2$ Density | $(3.11 \times 10^7 \text{ cm}^{-3}/R_{15}^2) \times$ $(\dot{M} / 3 \times 10^{-5} \text{ M}_\odot \text{ yr}^{-1}) \times (14.5 \text{ km s}^{-1}/v_r)$ where $R_{15}$ = radial distance/$10^{15}$ cm | 13 |
| Gas temperature | Max $[10, 12(90/R_{15})^{0.72}]$ K | 13 |
| Dust temperature | $1300 \, (R/R_i)^{-0.4}$ K | |
| Opacity due to dust | 0.5 cm$^2$ (wavelength / 50 $\mu$m)$^{-1.3}$ per gram of **gas** | |
| Assumed distance | 170 pc | 12 |

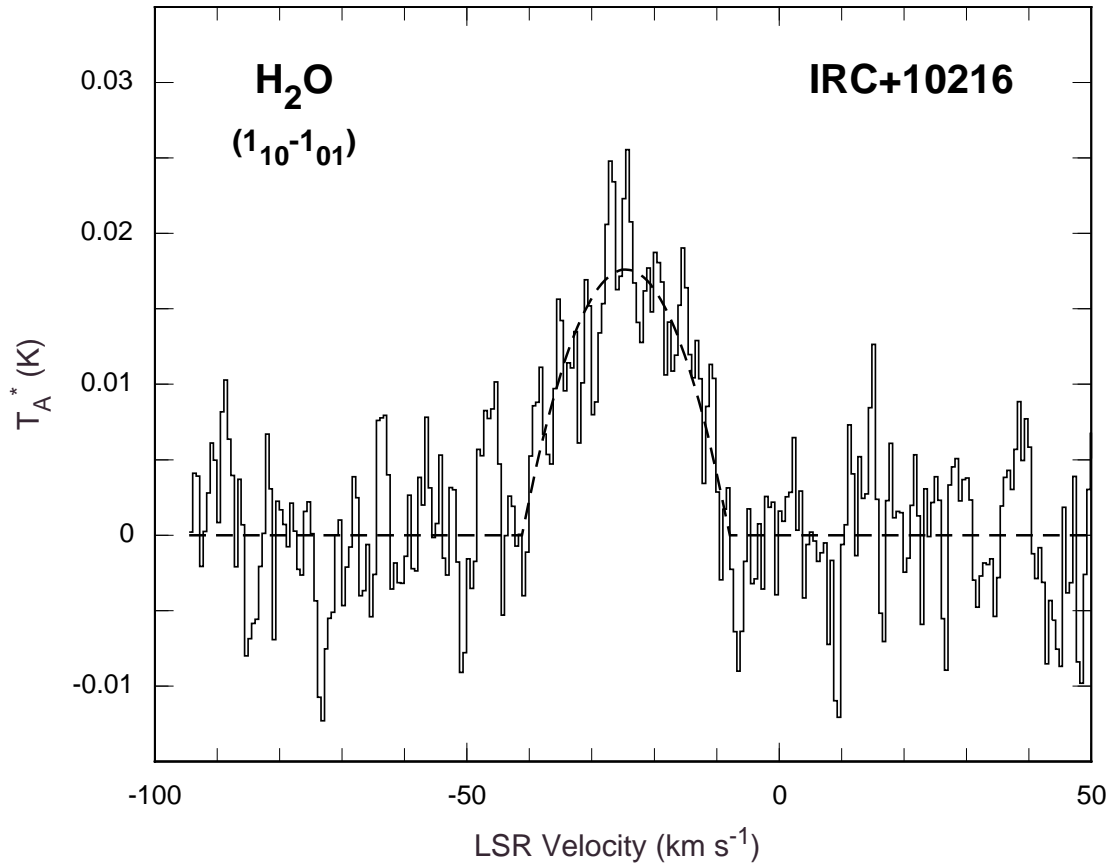

melnick_fig1. The *SWAS* $1_{10} - 1_{01}$ 556.936 GHz continuum-subtracted spectrum obtained toward IRC+10216. The dashed line is a parabolic curve fitted to the spectrum (see text). All data were obtained with the telescope pointed at the position $\alpha = 9^h \ 47^m \ 57.4^s$, $\delta = 13° \ 16' \ 44''$ (J2000). The observations were carried out by nodding the observatory between the source position and a reference position 30 arcminutes north of the source, selected to coincide with a region of no detectable molecular emission. The *SWAS* field-of-view at 557 GHz is 3.3 x 4.5 arcminutes. A total of 196.4 hours of on-source integration time has been obtained using three different local oscillator settings. The water line was evident in the signal sideband at all three settings. The data were reduced using the standard *SWAS* pipeline. Analyses of the spectral noise at frequencies separated from the expected line confirm that the system performs in accordance with radiometer theory (i.e., the signal-to-noise ratio improves $\propto \sqrt{\text{time}}$) and that the noise obeys Gaussian statistics[6].

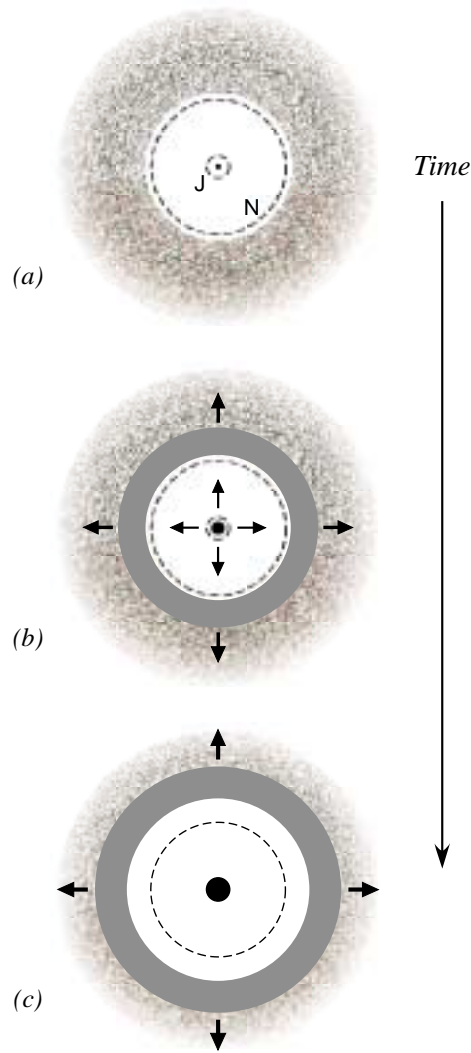

melnick_fig2.  Schematic drawing of IRC+10216 as it evolves from its main sequence to its asymptotic giant branch phase accompanied by an increase in both its size and luminosity, causing a wave of evaporation to propagate through the orbiting icy bodies.  Specifically: *(a)* during its main sequence phase, the luminosity of IRC+10216 is insufficient to vaporize icy bodies orbiting more distant than a few AU – for scale, distances equal to the orbits of Jupiter (J) and Neptune (N) are shown as dashed circles; *(b)* with the hydrogen and helium depleted within its core, IRC+10216 begins to expand, its luminosity increases, and a zone of vaporization (solid grey) begins to move outward through the population of icy bodies, depositing water vapor into the outflow; *(c)* IRC+10216's diameter has increased to a size roughly equivalent to the orbit of Jupiter, its luminosity has increased greatly, and the zone of vaporization expands. Detailed models by Ford and Neufeld (2001, in preparation) describe the evolution of a collection of icy bodies with distributions in size and orbital semi-major axis that are similar to those characterizing the Solar System's Kuiper Belt[17]. They indicate that water abundances as large as $\sim 10^{-7}$ $(M_0/M_\oplus)$ can be achieved in the outflow, where $M_0$ is the mass of orbiting water-ice at the onset of the post-main-sequence evolution and $M_\oplus$ is the mass of the Earth. This result is reproduced by a simple argument given above in the text.